%% file: funcons.tex
\documentclass[runningheads,fleqn]{llncs}
\usepackage{amsmath}
\usepackage{amssymb}

\usepackage{cbs-latex}
\usepackage[hyphens]{url}

\usepackage[breaklinks=true]{hyperref}
\usepackage{breakurl}

\newcommand{\QUESTION}[1]
{\pagebreak[3]\begin{itemize}\item \emph{#1} \end{itemize}}

\begin{document}

\title{Fundamental~Constructs in Programming~Languages}

\titlerunning{Fundamental Constructs}

\author{Peter~D.~Mosses\inst{1,2}\orcidID{0000-0002-5826-7520}}

\institute{Delft University of Technology, The Netherlands
\and Swansea University, United Kingdom
\\
\email{p.d.mosses@swansea.ac.uk}
}

\maketitle

\begin{abstract}
When a new programming language appears,
the syntax and intended behaviour of its programs need to be specified.
The behaviour of each language construct can be concisely specified
by translating it to fundamental constructs (funcons), compositionally.
In contrast to the informal explanations
commonly found in reference manuals,
such formal specifications of translations to funcons can be precise and complete.
They are also easy to write and read,
and to update when the language evolves.

The \textsc{PLanCompS} project has developed a large collection of funcons.
Each funcon is defined independently,
using a modular variant of structural operational semantics. 
The definitions are available online,
along with tools for generating funcon interpreters from them.

This paper introduces and motivates funcons.
It illustrates translation of language constructs to funcons,
and funcon definition.
It also relates funcons to the notation used in some previous language specification frameworks,
including monadic semantics and action semantics.

\keywords{
Funcons
\and
Programming constructs
\and
Formal specification
}
\end{abstract}


\section{Introduction}
\label{sec:introduction}

Many constructs found in (high-level) programming languages combine several behavioural features.
For example, call-by-value parameter passing in an imperative language involves order of evaluation,
allocating storage and initialising its contents, local name binding, and lexical scoping.
Such language constructs generally provide conciseness and clarity in programs,
and may support efficient implementation techniques (e.g., stack-based storage);
but their full behaviour can be quite difficult to understand, and tedious to specify directly.

Moreover, constructs in different languages may look the same but have very different behaviour
(e.g., the notorious `\verb"x=y"'),
or look different but have exactly the same behaviour
(e.g., `\verb"while...do..."' and `\verb"while(...){...}"').
Relatively minor differences between similar language constructs in different languages
include order of evaluation in expressions, and the effect of arithmetic overflow.
The evolution of programming languages has resulted in a huge diversity of language constructs
and their variants.
Some of the constructs are quite simple, but no programming \emph{lingua franca} has emerged.

The \textsc{PLanCompS} project%
\footnote{\url{https://plancomps.github.io}} 
has developed a large collection of \textit{fundamental constructs} (`funcons')
from which the behaviour of many high-level programming language constructs can be composed \cite{CBS-beta}.
The behaviour of a complete programming language can be concisely specified
by translating all its constructs, compositionally, to funcons.
In contrast to the informal explanations commonly found in language reference manuals,
formal specifications of translations to funcons can be precise and complete.
They are also easy to write and read,
and to update when the language evolves.
This could make them especially useful during design and development of domain-specific languages.

Funcons are significantly simpler than typical language constructs, in general:
\begin{itemize}

\item 
Each funcon affects only a single behavioural feature,
such as flow of control or data, name binding, storing, or interacting. 

\item
Variants of funcon behaviour (e.g., evaluating their arguments in a different order)
can be expressed by composite funcon terms.

\item
The funcons abstract from details related to implementation efficiency.

\end{itemize}

Funcon behaviour is defined using a modular variant \cite{MSOS,Mosses2009IPSOS}
of small-step structural operational semantics~\cite{SOS},
based on value-computation transition systems \cite{FOSSACS2013}.
Any program behaviour that can be modelled by a labelled transition system can, in principle,
also be specified by composing appropriately-defined funcons.
Thus specification by translation to funcons does not, per se, restrict the features of specified languages.
When the translation of a particular language construct is excessively complicated,
however, its design may be questionable.

The definition of a funcon determines its \textit{name}, its \textit{signature}, and its \textit{behaviour}.
Each funcon name should have a \emph{unique} definition, 
so that it always refers to the same signature and behaviour,
regardless of where the reference occurs.
To support reuse, funcon definitions need to be \emph{fixed} and \emph{permanent}:
changing or removing funcons would undermine the validity of translations that use them.
In particular, adding a new funcon to a collection
must never require changes to the definitions already in it.

Version control is superfluous for funcons;
translations of language constructs to funcons, in contrast, may need to change
when the specified language evolves.
For example, the illustrative language \textsc{Imp} includes a plain old while-loop
with a Boolean-valued condition:
`$\texttt{while} \texttt{(} \VAR{BExp} \texttt{)} \VAR{Block}$'.
The following rule translates it to the funcon $\NAME{while-true}$,
which has exactly the required behaviour:
\begin{align*}
  \KEY{Rule} \quad
    & \SEM{execute} \LEFTPHRASE \
        \LEX{while} \ \LEX{(} \ \VAR{BExp} \ \LEX{)} \ \VAR{Block} \ \RIGHTPHRASE  = \\&\quad
      \NAME{while-true}
        (  \SEM{eval-bool} \LEFTPHRASE \ \VAR{BExp} \ \RIGHTPHRASE , 
               \SEM{execute} \LEFTPHRASE \ \VAR{Block} \ \RIGHTPHRASE  )
\end{align*}
The behaviour of the funcon $\NAME{while-true}$ is fixed.
But suppose the \textsc{Imp} language evolves,
and a $\VAR{Block}$ can now execute a statement `$\texttt{break;}$',
which is supposed to terminate just the \textit{closest} enclosing while-loop.
We can extend the translation with the following rule:
\begin{align*}
  \KEY{Rule} \quad
    & \SEM{execute} \LEFTPHRASE \ \LEX{break} \ \LEX{;} \ \RIGHTPHRASE  = 
      \NAME{abrupt}(\NAME{broken})
\end{align*}
The translation of `$\texttt{while(true)\{break;\}}$' is 
$\NAME{while-true}(\NAME{true}, \NAME{abrupt}(\NAME{broken}))$.
The funcon $\NAME{abrupt}(V)$ terminates execution abruptly,
signalling its argument value~$V$ as the reason for termination.
However, the behaviour of $\NAME{while-true}(\NAME{true}, X)$ is to terminate abruptly whenever $X$ does
-- so this translation would lead to abrupt termination of \emph{all} enclosing while-loops!

We cannot change the definition of $\NAME{while-true}$,
so we are forced to change the translation rule.
The following updated translation rule reflects the extension of the behaviour of while-loops
with the intended handling of abrupt termination due to break-statements,
and that they propagate abrupt termination for any other reason:
\begin{align*}
  \KEY{Rule} \quad
    & \SEM{execute} \LEFTPHRASE \
        \LEX{while} \ \LEX{(} \ \VAR{BExp} \ \LEX{)} \ \VAR{Block} \ \RIGHTPHRASE  = \\&\quad
      \NAME{handle-abrupt}
        (  \\&\quad\quad
        \NAME{while-true}
            (  \SEM{eval-bool} \LEFTPHRASE \ \VAR{BExp} \ \RIGHTPHRASE , 
               \SEM{execute} \LEFTPHRASE \ \VAR{Block} \ \RIGHTPHRASE  ),\\&\quad\quad
            \NAME{if-true-else}
            (   \NAME{is-equal} ( \NAME{given}, \NAME{broken}), 
               \NAME{null-value},
               \NAME{abrupt}(\NAME{given})))
\end{align*}
Computing $\NAME{null-value}$ represents normal termination;
$\NAME{given}$ refers to the reason for the abrupt termination.

The specialised funcon $\NAME{handle-break}$ can be used
to specify the same behaviour more concisely:
\begin{align*}
  \KEY{Rule} \quad
    & \SEM{execute} \LEFTPHRASE \
        \LEX{while} \ \LEX{(} \ \VAR{BExp} \ \LEX{)} \ \VAR{Block} \ \RIGHTPHRASE  = \\&\quad
      \NAME{handle-break}
        (  
        \NAME{while-true}
            (  \SEM{eval-bool} \LEFTPHRASE \ \VAR{BExp} \ \RIGHTPHRASE , 
               \SEM{execute} \LEFTPHRASE \ \VAR{Block} \ \RIGHTPHRASE  ))
\end{align*}
Wrapping $ \SEM{execute} \LEFTPHRASE \ \VAR{Block} \ \RIGHTPHRASE$ in $\NAME{handle-continue}$ 
would also support abrupt termination of the current \emph{iteration} due to executing a continue-statement.

\paragraph{Overview.}

The reader is assumed to be interested in programming languages,
and familiar with their main concepts.
The research on which this paper is based has been published elsewhere
\cite{JLAMP,FOSSACS2013,TAOSD,MSOS,Mosses2019CBFLW,JVLC,Mosses2009IPSOS}.
The main aims here are to motivate the general idea of funcons,
and illustrate how they can be used to specify the behaviour of programming language constructs.

\begin{itemize}

\item
Section~\ref{sec:funcons} explains some general features of funcons.

\item
Section~\ref{sec:collections} considers how to manage large collections of funcons.

\item
Section~\ref{sec:facets} analyses various facets of funcon behaviour.

\item
Section~\ref{sec:languages} illustrates specification of translation of language constructs to funcons,
and explains how to validate such translations.

\item
Section~\ref{sec:msos} illustrates how to define funcons independently.

\item
Section~\ref{sec:related} relates funcons to the auxiliary operations defined in denotational semantics, 
to monads,
and to the combinators used in action semantics.

\item
Section~\ref{sec:conclusion} concludes with plans for future development of funcons,

\item Appendices~\ref{app:data}--\ref{app:interacting} give an informal summary of the currently defined funcons.%
\footnote{At the time of writing, the collection has not yet been released, and could change.}

\end{itemize}

The rest of this paper is structured as responses to questions
that readers might ask about funcons.
The author welcomes further questions,
as well as comments on the given responses.


\section{The Nature of Funcons}
\label{sec:funcons}

Let us start by explaining some general features of funcons.

\QUESTION{What aspects of behaviour do funcons represent?}
Funcons abstract from details related to implementation efficiency, 
such as storage allocation algorithms and communication protocols.
They express implementation independent behaviour that arises when programs are executed.
They also express linguistic features on which that behaviour depends,
such as scopes of bindings.

\QUESTION{Can funcons be implemented efficiently?}
Funcons need to be executable, to support validation of language translations.
Their current implementation uses \textsc{Haskell} interpreters
generated directly from their definitions \cite{Funcon.Tools}.
The efficiency of evaluating funcon terms is adequate for running unit tests and typical test programs,
but not applications.
However, it should be possible to implement certain sets of funcons more efficiently,
e.g., using virtual machines that support just the required features,
or by optimised compilation of funcon terms to other languages.

\QUESTION{How complicated are funcons ?}
One might expect that funcons should be as simple as possible.
In fact the aim is for funcons to be not too complicated, but not \emph{too} simple -- just right!
In the physical sciences, molecules are characterised and understood
primarily in terms of chemical bonds between their constituent atoms,
and atoms are formed from protons, neutrons, and electrons;
protons and neutrons are themselves composed from sub-atomic particles, such as quarks.
To explain a molecule in terms of sub-atomic particles might be possible, but unhelpful.
Language constructs are somewhat analogous to molecules, and funcons to atoms.

Introducing a funcon that corresponds directly to a complicated language construct
would make the funcon analysis of that language construct trivial,
but a direct definition of the funcon behaviour would necessarily be complicated.
At the other extreme, taking pure function abstraction and application as the only funcons
would make analysis and specification of language constructs as complicated as in 
(pre-monadic) denotational semantics.

Funcons aim to be unbiased towards any family of languages.
Adding an intermediate layer of not-so-fundamental constructs
that are closely related to some particular language constructs is thus undesirable.
However, it is sometimes appropriate to define funcons that abbreviate particular compositions
of other funcons.
Section~\ref{sec:introduction} mentioned $\NAME{handle-break}$,
which handles abrupt termination caused only by $\NAME{abrupt}(\NAME{broken})$ in $X$;
it abbreviates $\NAME{handle-abrupt}(X, Y)$ where $Y$ involves an explicit test
whether a given signal is the value $\NAME{broken}$.
Similarly, the funcon $\NAME{allocate-initialised-variable}(T, V)$ abbreviates
the sequential composition of $\NAME{allocate-variable}(T)$ and $\NAME{initialise-variable}(\_\,, V)$.%

\QUESTION{Can funcons have alternative behaviours?}
No, never.
The behaviour of common \textit{language} constructs, such as assignment expressions,
often varies significantly between different languages.
For example, the order of evaluation of the two sides of an assignment expression 
is left to right in some languages, right to left in others, or may even be implementation-dependent;
and the result may be the target variable or the assigned value.
The \textit{funcon} for assignment needs to have a behaviour from which all those variations can be obtained
by composition with other funcons.

\QUESTION{Are funcons independent?}
Funcons are often independent, but not always.
For instance, the definition of the funcon $\NAME{while-true}$
specifies the reduction of $\NAME{while-true}(B, X)$ to
a term involving the funcons $\NAME{if-true-else}$ and $\NAME{sequential}$:
\begin{align*}
  \KEY{Funcon} \quad
  & \NAME{while-true}(
                       B :  \TO \NAME{booleans}, X :  \TO \NAME{null-type}) 
    :  \TO \NAME{null-type} \\&\quad
    \leadsto \NAME{if-true-else}
               (  B, 
                      \NAME{sequential}
                       (  X, 
                              \NAME{while-true}
                               (  B, 
                                      X ) ), 
                      \NAME{null-value} )
\end{align*}
Duplication of $B$ before starting to evaluate it is essential,
in case it needs to be re-evaluated after the execution of $X$.
We could introduce an auxiliary term constructor for that,
but it is simpler to make use of $\NAME{if-true-else}$ and $\NAME{sequential}$.

\QUESTION{Do features of funcons interact?}
No. Feature interactions in software development tend to arise
when requirement specifications are incomplete.
An example of feature interaction in \cite{Batory2011FIPC}
involves a flood prevention system that turns off the water supply,
and a sprinkler system that depends on that water;
the requirements regarding flood prevention 
had better include checking the safety of turning the water off\dots

The complete requirement for each funcon is to provide just the behaviour specified in its definition,
propagating all \textit{unmentioned} effects of evaluating its arguments.
The values of the arguments are required to be consistent with the types in the funcon signature,
but \textit{no} further requirements arise when combining funcons.

\QUESTION{Can funcons be used as a programming language?}
Composing funcons is similar to the original idea of \textsc{Unix}: 
plugging simple commands together to produce complex behaviour.%
\footnote{Nowadays, a \textsc{Unix} command often has a multitude of obscure options, 
documented in a manual `page' that fills many screens.}
Not-so-fundamental constructs could be defined as abbreviations
for frequently-needed funcon compositions;
a~coating of `syntactic sugar' would be needed
to avoid an unwelcome plethora of parentheses in larger funcon compositions.

The main drawback of programming directly with funcons 
would be the comparatively low efficiency of their current implementation,
which uses interpreters written in \textsc{Haskell}, generated directly from funcon definitions.

\QUESTION{Can funcons be higher-order?}
Funcons can represent higher-order functions as values, but funcons are not themselves higher-order:
they do not take (non-constant) funcons as arguments.
However, it is easy to define funcons for common idioms of higher-order programming
(maps, filters, folds, etc.).

\QUESTION{Do funcons have algebraic properties?}
Yes: many binary funcons are associative, with left and right units;
some are also commutative.
These properties hold for a notion of bisimulation for the value-computation transition systems
\cite{FOSSACS2013}
that provide the foundations for funcon definitions.
This bisimulation is preserved when new funcons are added.
Funcon terms are written as applicative expressions,
and associativity allows binary funcons to be extended to longer sequences of arguments.

\QUESTION{Can I use my favourite proof assistant to prove properties of funcons?}
Some years ago, the modular variant \cite{MSOS} of structural operational semantics used to define funcons
was implemented in the \textsc{Coq} proof assistant,
and modular proofs of some properties were carried out \cite{Madelener2011FCBS}.
In a related line of work \cite{Torrini2015RMD}, 
a~different method for modular proofs in \textsc{Coq} was developed.

Modular proofs depend only on the definitions of the funcons involved,
and remain sound when funcon definitions are combined.
In principle, they could be released together with the funcons.


\section{Collections of Funcons}
\label{sec:collections}

The current collection of funcons is called \textsc{Funcons-beta}.
It includes several hundred funcons.
Management of such a collection is non-trivial.

\QUESTION{How can we classify funcons?}
Most high-level programming languages distinguish syntactically between commands (a.k.a.\ statements), 
declarations, and expressions.
Commands may assign to variables; 
declarations bind names;
and expressions compute values.
However, such syntactic distinctions are not universal:
for instance, expressions sometimes subsume commands, and sequences of commands may include declarations.
Grammars for programming language syntax (abstract as well as concrete)
often introduce many further syntactic distinctions.
A~universal set of syntactic sorts that encompasses all programming languages is not available.

For funcons, we have a single syntactic sort of \textit{terms}, with \textit{values} as a subsort.
A funcon term is similar to an expression in an (impure) functional programming language:
it computes values of a specific (possibly generic) type.
A funcon term corresponding to a command computes a fixed null value,
and a term corresponding to a declaration computes a value environment, mapping names to values.

We may also classify funcons according to their effects.
The behaviour of many funcons involves \textit{auxiliary entities}, representing various kinds of effects.
For instance, funcons for name binding use an auxiliary environment entity to represent the current bindings;
funcons for imperative variables use an auxiliary store entity
to represent the currently assigned values of variables.

\QUESTION{How do we refer to a particular funcon in a collection?}
A collection of funcons is like an \textit{open} package:
the names of all the funcons are visible externally (except those marked as auxiliary).
Neither the classification of funcons nor the paths to their definitions 
affects references to funcon names.

The name of each funcon should clearly suggest its behaviour,
to support casual reading of funcon terms
and the potential use of funcons as a controlled vocabulary
for informal discussion and comparison of programming languages.
Type names are \textit{plural} words (e.g., $\NAME{lists}$).%
\footnote{Singular forms of type names are used as value constructors.}
When a funcon corresponds directly to a familiar concept, a single well-chosen word can be adequate,
but otherwise several words (joined by hyphens) may be needed.
Moreover, different datatypes may have closely related operations,
yet the names for the corresponding funcons have to be distinct,
due to the absence of overloading:
the name of the datatype can be added as a prefix of the name, e.g., $\NAME{integer-add}$.%
\footnote{Currently, \textsc{Funcons-beta} does not support namespaces in collections of funcons.}

Suggestive names can be quite long,
and abbreviations may be needed in some situations
(e.g., classrooms, examinations, presentations).
Abbreviations can be defined as explicit aliases for funcons;
for instance, $\NAME{alloc-init}$ is defined as an alias for $\NAME{allocate-initialised-variable}$.

\QUESTION{Do funcons evolve?}
After a collection of funcons has been released,
the behaviour of all the funcons in it needs to be \textit{fixed} and \textit{permanent},
since changes could affect or break their uses in language translations
(which do not need to be public or registered).
All uses of a particular funcon name thus refer to the same behaviour.

However, the collection itself \textit{can} evolve: by extension with new funcons.
This must not require changes to the definitions of the previous funcons.
New funcons need to be carefully checked and tested before they are added,
since their definitions cannot be revoked.

Names of funcons always refer implicitly to the current version of a collection.
Evolving collections of funcons have no need for version \textit{numbers},
since once a funcon has been defined, adding definitions for new funcons
(or an alias for an already defined funcon)
cannot invalidate existing references to names.

\QUESTION{Will the \textsc{Funcons-beta} collection of funcons ever be finalised?}
\textsc{Funcons-beta} is a release \emph{candidate}.
After further polishing, review, and use in language specifications,
the collection of funcons and their documentation are to be released for general use.
However, it will always be possible to add new funcons to the collection,
so as to support new concepts or provide alternative ways of expressing existing concepts.


\section{Facets of Funcons}
\label{sec:facets}

When funcon terms are evaluated, their behaviour may have many \textit{facets}:
apart from computing values, funcon behaviour can involve
name bindings, imperative variables, abrupt termination, interaction, etc.
Facets that are not needed for a particular term can be ignored.

In this section, we introduce the main facets of the \textsc{Funcons-beta} collection.
Appendices~\ref{app:data}--\ref{app:interacting} provide an informal summary of the funcons;
their definitions are available online \cite{CBS-beta}.

\QUESTION{How are funcon terms evaluated?}
Evaluation of a funcon term may terminate \textit{normally},
\textit{abruptly}, or \textit{never}.
The evaluation takes a sequence of argument \textit{terms};
on normal termination, it computes a sequence of \textit{values}
(where a sequence of length 1 is identified with its only element).
The funcon signature specifies how many arguments it takes,
the type of values to be computed by each argument, and the types of values that the funcon computes.
Individual arguments may be required to be \textit{pre-computed} values;
the funcon definition specifies how its behaviour combines the computations of any remaining arguments.

\QUESTION{Does each funcon take a fixed number of arguments?}
Not necessarily: a funcon signature can specify that an argument at some position is optional,
or that it can be a sequence.
Sequence arguments are often used to extend associative binary funcons to longer argument sequences.
They are also used for funcons that correspond directly
to conventional notation for (finite) lists and sets, 
e.g., $\NAME{list}(V_1,\dots,V_n)$ for $[V_1,\dots,V_n]$.

\QUESTION{How do funcons represent data?}
Data that programs process when executed is represented by funcon terms classified as values.
Some funcons are value constructors:
they are \textit{inert}, and have no computational behaviour themselves.
Values are themselves classified as \textit{primitive values}, 
\textit{composite values}, or (procedural) \textit{abstractions}.

Conceptually, primitive values are atomic, and not regarded as constructed from other values.
Booleans, unbounded integers, IEEE floats, \textsc{Unicode} characters,
and a null value are all classified as primitive.
Some of them have constant constructors;
the rest are computed by built-in funcons.

Composite values are constructed from finite sequences of argument values.
Value constructors are injective: different argument value sequences give different composite values.
The types of composite values include parametrised algebraic data types, with a generic representation.
Various algebraic datatypes are predefined, and new ones can be introduced.
Composite values include also built-in parametrised types of sets, maps, multi-sets, and graphs.

Abstractions are values formed by the value constructor $\NAME{abstraction}(X)$
with an \textit{unevaluated} argument~$X$. 
Values are called \textit{ground} when they are constructed entirely from primitive and composite values,
without any abstraction values.

Appendix~\ref{app:data} summarises the funcons for types of data,
and some funcons for data operations.

\QUESTION{What kind of behaviour do funcons for data operations have?}
Data operations in programs are generally represented by funcons
whose only behaviour is to compute values from \textit{pre-evaluated} arguments.
The arguments are evaluated in \textit{any} order, possibly with interleaving
(the order of argument evaluation is irrelevant when the evaluations have no effects).
\textit{Partial} data operations (e.g., integer division, or selecting the head of a list)
compute the empty sequence when their arguments are not in their domain of definition.

Value \textit{types} are themselves values,
so funcons can take types as arguments and give them as results.
Apart from supporting dependent types, this generality is needed 
to represent ordinary type constructors as funcons
(e.g., $\NAME{lists}(T)$,
where~$T$ is the type of the list elements).

\QUESTION{How do funcons express normal flow of control?}
A funcon intended purely for specifying normal control flow
generally specifies the potential order of evaluation of its arguments,
but does not otherwise contribute to behaviour.
Such funcons include sequential or interleaved command execution and expression evaluation,
deterministic and non-deterministic choice between computations,
and command iteration.

Appendix~\ref{app:controlflow} summarises the funcons for representing control flow.

\QUESTION{How do funcons express flow of data?}
A computation may involve multiple uses of the same data
(e.g., so as to assign it to a variable as well as provide it as a result).
It may also involve repeating the same computation with different data.
The computations of funcons for specifying such data flow involve
an auxiliary entity $\NAME{given-value}(V)$
that can be set to a computed value~$V$;
the funcon $\NAME{given}$ gets the current value.

Appendix~\ref{app:dataflow} summarises the funcons for representing data flow.

\QUESTION{How do funcons specify scopes of  bindings?}
An occurrence of a name in a program either \textit{binds} the name,
or \textit{references} whatever is currently bound to the name.
Binding occurrences are usually found in declarations, parameter specifications, and patterns;
references to names are ubiquitous.
Sequences of declarations have the effect of successively extending (or perhaps overriding)
the current bindings with the bindings due to the individual declarations.

Funcons use conventional environments~$\rho$ (mapping names to values)
to represent both the current bindings
and bindings computed by declarations.
The auxiliary entity $\NAME{environment}(\rho)$ represents the current bindings;
the current binding for an individual name~$I$ is inspected using the funcon $\NAME{bound-value}(I)$.
An environment representing computed bindings is an ordinary composite value,
and can be inspected using data operations.

Some languages include various constructs for composing declarations,
and these are represented directly by funcons that compute environments.
However, the funcons corresponding to recursive declarations represent circularity
by creating cut-points called links,
which involves a separate auxiliary entity.

Appendix~\ref{app:binding} summarises the funcons for representing name binding.

\QUESTION{Do funcons have static scopes for bindings?}
The difference between static and dynamic scopes concerns procedural abstraction.
A value that represents an abstraction is constructed from an \textit{unevaluated} argument~$X$
by the funcon $\NAME{abstraction}(X)$. 
The abstraction value can be subsequently enacted, which evaluates the argument~$X$ 
-- potentially in a different context from that where the abstraction value was constructed.

Constructed abstraction values thus naturally have \textit{dynamic} scopes for bindings. 
To obtain \textit{static} scopes, the funcon $\NAME{closure}(X)$ computes a closure value:
an abstraction whose argument evaluation starts by ignoring the current bindings
and (locally) re-declaring the abstraction-time bindings. 

\QUESTION{How do funcons distinguish between constant and mutable variables?}
In programming languages, imperative variables usually have names.
It may be tempting to regard variable names as \textit{bound} directly to values:
bindings then need to be mutable, assignment to a variable name updates its binding,
and constants correspond to single-assignment variables.
However, such a simplistic analysis does not easily extend
to features such as aliasing and call by reference.

A more satisfactory conceptual basis for imperative variables is 
to regard them as independent storage \textit{locations}.%
\footnote{Funcons have not yet been developed for `relaxed' memory models or data marshalling.}
The declaration of a named variable involves allocation of storage
(optionally with an initial value)
together with binding the name of the variable to the storage location.
Assignment to a named variable then affects what value is stored at the location,
but leaves the bindings unchanged.
Aliasing can now be understood simply as the simultaneous binding of different names to the same location.

The funcons for imperative variables involve an auxiliary mutable entity $\NAME{store}(\sigma)$,
mapping locations to their currently assigned values.
The store supports allocation (and recycling) of locations for values of any type,
and their initialisation, assignment, and inspection.
It is completely independent of the auxiliary entity $\NAME{environment}(\rho)$
used to represent the current name bindings.

In mathematical logic, a `variable' corresponds to a name,
and `assignment' to binding.
Imperative variables in programming languages are often called `L-values',
with `R-values' being those that can be assigned to variables.%
\footnote{
`L' and `R' refer to the left and right sides of typical assignment commands \cite{Strachey1967FCPL}.}
With funcons, all values can be assigned to variables -- and variables are themselves values.

\QUESTION{Can funcons represent structured variables with mixtures of constant and mutable fields?}
A simple variable consists of a location together with the type of values that it can store;
assignment checks that the value to be assigned to the variable is in its type.%
\footnote{Funcons for using un-typed locations as variables would be slightly simpler.}
Simple variables may store primitive values (e.g., numbers) or composite values (e.g., tuples),
but assignment to a simple variable is always \textit{monolithic}:
the current value is replaced \textit{entirely} by the new value.

Structured variables are composite values where some components are simple variables.
These include hybrids having both mutable and immutable components.
Assignment to a selected component variable corresponds to an in-place update;
assignment of a composite value to an entire structured variable
updates all the component simple variables with the matching values,
and checks that the immutable components are the same.

Appendix~\ref{app:storing} summarises the funcons for representing imperative variables.

\QUESTION{How about abrupt termination?}
Various language constructs may cause abrupt termination when executed:
throwing or raising an exception, returning the value of a function,
breaking out of a loop, etc.
Enclosing constructs can detect particular kinds of abrupt termination,
and handle them appropriately.
For example, a language construct may inspect an exception value,
and conditionally handle it;
a function application handles an abruptly returned value by giving it as the result;
and a loop handles a break by terminating normally.

Funcons express abrupt termination and handlers \textit{uniformly}.
Evaluation of a funcon term may terminate normally, abruptly, or never.
Abrupt termination leads to a stuck term,
emitting an auxiliary entity $\NAME{abrupted}(V)$ as a \textit{signal} with a value~$V$.
The \textit{closest} enclosing funcon that notices the emission of such a signal
can inspect its value, and determine whether to handle it or not.

Appendix~\ref{app:abrupting} summarises the funcons for abrupt termination.

\QUESTION{Is it possible to define delimited control operators as funcons?}
Somewhat surprisingly, yes: see \cite{CBS-beta,Sculthorpe2016MSOSDC}.
Control operators include continuation handling functions, such as `\texttt{call-cc}'.

\QUESTION{Can non-terminating funcon evaluation have observable behaviour?}
Yes: through \textit{interactive} input and output.

Program behaviour may depend on, and affect, data stored in files.
Conceptually, files can be regarded as (complicated) structured variables:
input from a file inspects the value stored at the current position, and advances the position;
output to a file appends a value to it.
Changes to a file system during program execution correspond to updating values stored in locations;
they may subsequently be overwritten,
so their \textit{final} values can only be observed on program termination.

Interactive input and output, in contrast, \textit{cannot} be regarded as effects on mutable storage.
Acceptance of input data from a stream during program execution is irrevocable,
as is output of data to a stream.
Interaction may also involve inter-dependence between input and output.
And a program that never terminates can have infinitely long streams of input and output.

Thus funcons for expressing interaction involve kinds of entities
that differ fundamentally from those we previously introduced.
The auxiliary entity $\NAME{standard-in}(V^*)$ 
represents the (finite) sequence of values \textit{input} at each step of a computation,
where the empty sequence $(   \  )$ represents that no values are input.
The value $\NAME{null-value}$ indicates the end of the input.
The auxiliary entity $\NAME{standard-out}(V^*)$
represents the (finite) sequence of values \textit{output} at a particular step,
where the empty sequence $(   \  )$ represents the lack of output.
Computations concatenate the input sequences of each step, and similarly for output --
potentially resulting in infinite sequences for non-terminating computations.

Appendix~\ref{app:interacting} summarises the funcons for representing interaction.
To support multiple streams, further entities and funcons would need to be added.

\QUESTION{Do funcons currently support specification of any other language features?}
Tentative funcons for \textit{multithreading} have been developed.
They have not yet been rigorously unit-tested, nor used much in language definitions.
These funcons are not included in \textsc{Funcons-beta},
but in an unstable collection that extends \textsc{Funcons-beta}
\cite{CBS-beta}.

The multithreading funcons involve multiple mutable auxiliary entities,
representing the collection of threads, the set of active threads,
the thread being executed,
the values computed by terminated threads,
and (abstract) scheduling information.
Funcons that combine effects on multiple auxiliary entities are undesirable,
and their definitions are somewhat verbose.
It is currently unclear whether simpler funcons for multithreading can be developed.

Multithreading also involves \textit{synchronisation}.
The funcons for synchronising involve only the store entity.
To inhibit preemption during synchronisation,
multiple assignments need to be executed atomically, in a single transition.

Funcons for \textit{distributed processes} have not yet been developed.
They are expected to be based on asynchronous execution and message passing
(cf.\ \cite{Mosses1992AS}).

Funcons for specifying \textit{meta-programming} constructs have been defined \cite{Binsbergen2018FHGMP};
they also enable a straightforward specification of call-by-need parameters.


\section{Translation of Language Constructs to Funcons}
\label{sec:languages}

In this section, we illustrate how a simple programming language construct
can be specified by translation to funcons.
Specifying such a translation for all constructs of a language defines the behaviour of programs,
based on the behaviour of the funcons used in the translation.
The \textsc{PLanCompS} project has developed some examples \cite{CBS-beta}
and made them available for browsing on a website.
We conclude this section with an overview of the examples,
and indicate how they have been developed and tested.

\QUESTION{How is call-by-value translated to funcons?}
The following fragments of a language specification illustrate how call-by-value parameter passing
in an imperative programming language can be 
specified
by translation to funcons.
The fragments originate from a published specification \cite{TAOSD}
of the \textsc{Simple} language;
for brevity, however, we here restrict \textsc{Simple} function applications and declarations
to a single parameter.

The translation specification in Fig.~\ref{fig:expressions} declares $\SYN{exp}$ as a phrase sort,
with the meta-variable $\VAR{Exp}$ (possibly with subscripts and/or primes)
ranging over phrases of that sort.
The BNF-like production shows two language constructs of sort $\SYN{exp}$:
an identifier of sort $\SYN{id}$
(lexical tokens, here assumed to be specified elsewhere with meta-variable $\VAR{Id}$)
and a function application written `$\VAR{Exp}_1 \texttt{(} \VAR{Exp}_2\texttt{)}$'.

\begin{figure}
    \centering
\begin{align*}
  \KEY{Syntax} \quad
    \VAR{Exp} : \SYN{exp}
      ::=  \cdots 
      \mid \SYN{id}
      \mid \SYN{exp} \ \LEX{(} \ \SYN{exp} \ \LEX{)}
      \mid \cdots
\end{align*}
\begin{align*}
  \KEY{Semantics} \quad
  & \SEM{rval} \LEFTPHRASE \ \_ : \SYN{exp} \ \RIGHTPHRASE  
    :  \TO \NAME{values} 
\\
  \KEY{Rule} \quad
    & \SEM{rval} \LEFTPHRASE \ \VAR{Id} \  \RIGHTPHRASE  = 
      \NAME{assigned-value}
        (  \NAME{bound-value}
        ( \SEM{id} \LEFTPHRASE \ \VAR{Id} \ \RIGHTPHRASE  ) )
\\
  \KEY{Rule} \quad
    & \SEM{rval} \LEFTPHRASE \ \VAR{Exp}_1 \ \LEX{(} \ \VAR{Exp}_2 \ \LEX{)} \ \RIGHTPHRASE  = 
      \NAME{apply}
        (  \SEM{rval} \LEFTPHRASE \ \VAR{Exp}_1 \ \RIGHTPHRASE , 
               \SEM{rval} \LEFTPHRASE \ \VAR{Exp}_2 \ \RIGHTPHRASE  )
\end{align*}
    \caption{Translation of identifiers and function applications in \textsc{Simple} to funcons}
    \label{fig:expressions}
\end{figure}

The translation function $\SEM{rval}\LEFTPHRASE \VAR{Exp} \RIGHTPHRASE$
maps phrases $\VAR{Exp}$ of sort $\SYN{exp}$ to funcon terms that compute elements of type $\NAME{values}$.
Translation is compositional:
the funcon term for a phrase combines the translations of its sub-phrases.
The translation function $\SEM{id}\LEFTPHRASE \VAR{Id} \RIGHTPHRASE$ 
maps lexical tokens $\VAR{Id}$ of sort $\SYN{id}$ to funcon values of type $\NAME{identifiers}$
(its specification is omitted here).

In this illustrative language, the only values to which identifiers can be bound
are simple imperative variables.
When identifiers can be bound directly to other values (e.g., numbers)
we would use $\NAME{current-value}$ instead of $\NAME{assigned-value}$.

For call-by-value parameters in an imperative language,
the argument value can be passed to the called function,
which then has to allocate a variable to store the value.
For call-by-reference, the argument would have to evaluate to a variable;
for call-by-name, the evaluation of the argument would be deferred,
which can be expressed by constructing a thunk abstraction value from it.
When the mode of parameter-passing in function applications depends on the function,
argument evaluation needs to be incorporated in the value that represents the function.

The translation specification for function declarations in Fig.~\ref{fig:declarations}
assumes a translation function $\SEM{exec}\LEFTPHRASE \VAR{Block} \RIGHTPHRASE$ 
for phrases $\VAR{Block}$ of sort $\SYN{block}$.
A block is a statement, which normally computes a null value;
but here, as in many languages, a block can return an expression value by executing a return statement,
which terminates the execution of the block abruptly.

\begin{figure}
    \centering

\begin{align*}
  \KEY{Syntax} \quad
    \VAR{Decl} : \SYN{decl}
      ::= \cdots \mid 
      \LEX{function} \ \SYN{id} \ \LEX{(} \ \SYN{id} \ \LEX{)} \ \SYN{block}
\end{align*}
\begin{align*}
  \KEY{Semantics} \quad
  & \SEM{declare} \LEFTPHRASE \ \_ : \SYN{decl} \ \RIGHTPHRASE  
    :  \TO \NAME{environments} 
\\[1ex]
  \KEY{Rule} \quad
    & \SEM{declare} \LEFTPHRASE \
        \LEX{function} \ \VAR{Id}_1 \ \LEX{(} \ \VAR{Id}_2 \ \LEX{)} \ \VAR{Block} \
                          \RIGHTPHRASE  = \\&\quad
      \NAME{bind-value}
        ( \SEM{id} \LEFTPHRASE \ \VAR{Id}_1 \ \RIGHTPHRASE , \\&\quad\quad
               \NAME{allocate-initialised-variable}
                (  \NAME{functions}
                        (  \NAME{values}, 
                           \NAME{values} ),
                    \\&\quad\quad\quad
               \NAME{function} 
                ( \NAME{closure}
                  ( \\&\quad\quad\quad\quad \NAME{scope}
                          ( \\&\quad\quad\quad\quad\quad \NAME{bind-value}
                                  (  \SEM{id} \LEFTPHRASE \ \VAR{Id}_2 \ \RIGHTPHRASE ,\\&\quad\quad\quad\quad\quad\quad
                            \NAME{allocate-initialised-variable}
                (  \NAME{values}, 
                                  \NAME{given} ) ), \\&\quad\quad\quad\quad\quad
                                 \NAME{handle-return}
                                  (  \SEM{exec} \LEFTPHRASE \ \VAR{Block} \ \RIGHTPHRASE  ) ) ) ) )
                    )
\end{align*}

    \caption{Translation of function declarations in \textsc{Simple} to funcons}
    \label{fig:declarations}
\end{figure}

The use of $\NAME{closure}$ ensures static (lexical) bindings for references to names in the function body.
For dynamic bindings, we would replace $\NAME{closure}$ by $\NAME{abstraction}$.
The construction of a function value from the closure is needed 
so that $\NAME{apply}$ can be used to give the argument value to the body of the abstraction.

The $\NAME{scope}$ funcon adds the bindings computed by its first argument to the current bindings
for the evaluation of its second argument.
In this simplified illustration, functions have only one formal parameter,
which is bound to a freshly allocated variable containing the given argument value;
for multiple parameters, the given value would be a tuple of the same length,
matched by a pattern tuple.

The $\NAME{handle-return}$ funcon concisely handles abrupt termination of the function body
arising from evaluation of the $\NAME{return}$ funcon.
It has no effect on normal termination, nor on abrupt termination for other reasons.

In languages where function identifiers can be bound directly to function closures,
the first $\NAME{allocate-initialised-variable}$ in the translation rule could be eliminated.
However, the possibility of recursive function calls would then need to be expressed directly,
using the $\NAME{recursive}$ funcon.

The call-by-value example illustrates how directly the behaviour of a language construct
can be specified by translation to funcons.

\QUESTION{Which other language constructs have been translated to funcons?}
The \textsc{PLanCompS} project has developed the following 
language specifications based on \textsc{Funcons-beta},
and made them available for browsing online \cite{CBS-beta}.

\begin{itemize}

\item \textsc{Imp}:
a very small imperative language, often used in text books on semantics.
Its translation to funcons illustrates basic features of the framework.

\item \textsc{Simple}:
a somewhat larger imperative language than \textsc{Imp}\@.
Its translation to funcons \cite{TAOSD} illustrates most features of the framework.
It is comparable to the specification of \textsc{Simple} in K \cite{K},
except that multithreading is omitted.

\item \textsc{MiniJava}:
a very simple subset of \textsc{Java}, used in \cite{Appel2002MCIJ}.
Its specification illustrates translation to funcons for classes and objects.

\item SL:
the \textsc{SimpleLanguage} used for demonstration of \textsc{GraalVM} \cite{GraalVM}. 
Its translation to funcons illustrates how dynamic bindings can be specified.

\item \textsc{OCaml Light}:
a core sublanguage of \textsc{OCaml}.
Its specification illustrates how translations to funcons scale up to a medium-sized language.

\end{itemize}
Further examples of language specifications involve funcons from an unstable collection 
that extends \textsc{Funcons-beta} \cite{CBS-beta}:

\begin{itemize}

\item \textsc{Imp++}:
extends \textsc{Imp} with multithreading and various other features.

\item \textsc{Simple-Threads}:
adds the previously-omitted multithreading constructs.

\item \textsc{LangDev-2019}:
demonstrates extensibility of language specifications.

\end{itemize}
A funcon-based specification of C$\sharp$ is currently being developed.

\QUESTION{How can we check translations of language constructs to funcons?}
Consider our translation of function declarations with call-by-value parameters.
Potential mistakes include spelling errors in names
(primarily funcons, but also syntax sorts, translation functions, and meta-variables)
and misplaced parentheses.
The syntax of the language construct in the translation rule
might not be consistent with the specified grammar.
A less obvious mistake is when the arguments of a funcon could compute values
that are not in the types required by the funcon signature.
We might also have used a funcon that does not have the intended behaviour
(e.g., using $\NAME{abstraction}$ instead of $\NAME{closure}$).

Clearly, tool support for checking is essential.
A workbench for specifying translations of languages to funcons has been developed \cite{Mosses2019CBFLW}.
Tools for evaluating funcon terms \cite{Funcon.Tools} allow us to check
whether they have the expected behaviour.

The workbench checks references to names, term formation,
and the syntax in translation rules.
It checks that funcons have the right number of arguments,
but not yet that the arguments compute values of the required types;
we currently rely on testing to check for that.

The workbench also supports parsing complete programs and translating them to funcon terms,
using parsers and translators generated from the specified grammar and translation rules.
It is based on the \textsc{Spoofax} language workbench \cite{Spoofax-2010},
and implemented using the declarative domain-specific meta-languages \linebreak 
\textsc{Sdf3}, \textsc{NaBL2}, and \textsc{Stratego}.
See \cite{Mosses2019CBFLW} for further details.
The tools for evaluating funcon terms \cite{Funcon.Tools} are implemented in \textsc{Haskell},
and can be called directly from the workbench.


\section{Defining and Implementing Funcons}
\label{sec:msos}

In this section, we illustrate how to define the behaviour of a funcon, once and for all, 
using a highly modular variant \cite{MSOS,Mosses2009IPSOS} of structural operational semantics \cite{SOS}.
Modularity of funcon definitions is crucial for extensibility of funcon collections.

\QUESTION{How are funcons defined?}
The funcon signature in Fig.~\ref{fig:scope} specifies that $\NAME{scope}$ takes two arguments.
The first argument is required to be pre-evaluated to a value of type $\NAME{environments}$;
the second argument should be unevaluated, as indicated by `$\TO T$'.
Values computed by $\NAME{scope}( \rho_1, X )$ are to have the same type ($T$)
as the values computed by $X$.
\begin{figure}
    \centering

\begin{align*}
  \KEY{Funcon} \quad
  & \NAME{scope}( \_ : \NAME{environments}, \_ :  \TO T) 
    :  \TO T 
\\
  \KEY{Rule} \quad
    & \RULE{
      \NAME{environment} (  \NAME{map-override}
                                     (  \rho_1, 
                                            \rho_0 ) ) \vdash X \TRANS 
          X'
      }{
      \NAME{environment} (  \rho_0 ) \vdash \NAME{scope}
                      (  \rho_1 : \NAME{environments}, 
                             X ) \TRANS 
          \NAME{scope}
            (  \rho_1, 
                   X' )
      }
\\
  \KEY{Rule} \quad
    & \NAME{scope}
        (  \_ : \NAME{environments}, V : T ) \leadsto V
\end{align*}

    \caption{Definition of the funcon for expressing scopes of local declrations}
    \label{fig:scope}
\end{figure}

The rules define how evaluation of $\NAME{scope}( \rho_1, X )$ can proceed
when the current bindings are represented by $\rho_0$.
The premise of the first rule holds if $X$ can make a transition to $X'$
when~$\rho_1$ overrides the current bindings~$\rho_0$.
Whether $X'$ is a computed value or an intermediate term is irrelevant.
When the premise holds, the conclusion is that $\NAME{scope}( \rho_1, X )$
can make a transition to $\NAME{scope}( \rho_1, X' )$.

If $X$ can terminate abruptly, or continue making transitions forever,
then $\NAME{scope}( \rho_1, X )$ can do the same.
The last rule allows evaluation of $\NAME{scope}( \rho_1, X )$ to terminate normally,
computing the same value $V$ as $X$.
Transitions written with `$\leadsto$' correspond to term rewriting \cite{FOSSACS2013},
and do not involve auxiliary entities.

\QUESTION{How can funcon definitions remain fixed when new funcons are added?}
The use of the auxiliary entity $\NAME{environment} (  \rho_0 )$ in the definition of $\NAME{scope}$
restricts transitions to states that include it,
but states might still include other auxiliary entities,
such as $\NAME{store}(\sigma)$ or $\NAME{given-value}(V)$.
If a transition $X \TRANS X'$ updates $\sigma$ to $\sigma'$,
so does
$\NAME{scope}( \rho_1, X ) \TRANS \NAME{scope}( \rho_1, X' )$;
the transitions in the premise and conclusion use the same given value;
and if $X \TRANS X'$ emits a signal on abrupt termination, so does the corresponding transition for
$\NAME{scope}( \rho_1, X ) $.

Auxiliary entities are classified according to how they are \textit{propagated}: 
\begin{description}

\item[Contextual:]
A contextual entity remains \textit{fixed} for successive steps in the computation of a term,
but can be different for the computations of sub-terms.

\item[Mutable:]
Sequential \textit{changes} to a mutable entity are propagated between the computation of a term and 
the computations of its sub-terms.

\item[Input:]
An input entity is a sequence of values \textit{consumed} by evaluating a term,
concatenating the sequences consumed by the computations of its sub-terms.

\item[Output:]
An output entity is a sequence of values \textit{produced} by evaluating a term,
concatenating the sequences produced by the computations of its sub-terms.

\item[Control:]
A control entity is a value that can optionally be \textit{signalled} by a step.
The corresponding step of an enclosing term may inspect the value,
and signal the same value, signal a different value, or not signal.

\end{description}
The notation used for specifying auxiliary entities determines their classification.
For instance, entities written before `$\vdash$' are classified as contextual.

\QUESTION{Can static semantics for funcons be defined in the same way as dynamic semantics?}
The modular structural operational semantics rules for funcon term evaluation are in 
the\textit{ small-step} style,
where each rule has \textit{at most one} transition premise.
A static semantics for funcons would naturally use \textit{big-step} rules,
with a premise for each sub-term.
It is currently unclear whether the same classification of entities can be used
for static and dynamic semantics;
the static semantics of abstractions generally requires making latent effects explicit,
in contrast to dynamic semantics.

\QUESTION{How have funcons been implemented?}
The initial implementation of funcons was in \textsc{Prolog}.
Funcon definitions were translated to \textsc{Prolog} clauses defining transitions,%
\footnote{\url{https://pdmosses.github.io/prolog-msos-tool}}
based on the original implementation of MSOS in \textsc{Prolog}.%
\footnote{\url{https://pdmosses.github.io/msos-in-prolog}}
Funcons have also been implemented in \textsc{Maude}.%
\footnote{\url{https://github.com/fcbr/mmt}}
The \textsc{Prolog} implementation of MSOS was subsequently enhanced to support the rewriting relation
used in value-computation transition systems \cite{FOSSACS2013}.
The \textsc{Funcon Tools} package \cite{Funcon.Tools} supports parsing funcon definitions 
and generating funcon interpreters in \textsc{Haskell}, as described in \cite{JLAMP}.

\QUESTION{Could funcons be used for language specification in other frameworks?}
The K-framework \cite{K} has a high degree of modularity.
For an experiment with using the K-framework to define funcons, see \cite{FunKons}.
The distinction between pre-evaluated and unevaluated arguments in funcon signatures
is represented by strictness annotations in K\@. 
However, rules in K are unconditional,
so funcons such as $\NAME{scope}$ cannot be defined straightforwardly.
The specification of the structure of states is monolithic,
and may need updating when adding new funcons.

\textsc{Redex} \cite{Klein2012RYR} is a popular domain-specific metalanguage for operational semantics, 
embedded in the \textsc{Racket} programming language.
It is based on reduction rules and evaluation contexts.
The reduction rules are highly modular,
and grammars for language constructs and evaluation contexts can be specified incrementally.
However, evaluation context grammars associated with control operators appear to be inherently global.
It should be possible to define a particular collection of funcons in \textsc{Redex},
but adding a new funcon could require updating the evaluation contexts for existing funcons.


\section{Related Work}
\label{sec:related}

Many funcons are closely related to notation used in several
previously developed language specification frameworks:
denotational semantics, monads, abstract semantic algebras, and action semantics.

\paragraph{Denotational Semantics.}

The funcons for flowing, binding, and storing are directly based on 
Christopher Strachey's original conceptual analysis of imperative programming languages.
Strachey initiated the development of denotational semantics
at the IFIP Working Conference on \emph{Formal Language Description Languages} in 1964 \cite{Strachey1966TFS}.
At the time, he was working on the design and implementation of the high-level CPL programming language,
and aiming to specify its semantics formally.
In the paper, he focuses on representing imperative features of programming languages
as pure mathematical functions, avoiding the introduction of abstract machines.
For assignment commands, he distinguishes between L-values and R-values of expressions,
with locations in stores $\sigma$ being a special case of L-values.
He defines the operation $C$  to get the current content of a location, and $U$ to update the content.
For flow of control, he uses composition of functions from stores to stores,
and the fixed-point operation $Y$.
In his widely-circulated 1967 lecture notes \cite{Strachey1967FCPL}, 
he also introduces environments that map names to values,
and represents procedures as closures.

Strachey's original operation  $C$ on stores is renamed \textit{Contents} in \cite{Scott1971TMSCL},
and $U$ is renamed \textit{Assign}.
Many subsequent denotational specifications define a large number of such auxiliary operations
(e.g., \cite{Mosses1974MSA} defines about 80).
However, the definitions are ad hoc, and they are based on the domains defined for the specified language.
Even the way lambda-expressions are written, and the notation used for modifying environments and stores,
vary between denotational specifications. 

The VDM metalanguage for denotational semantics, developed from 1974 \cite{Jones2001TVV},
introduced fixed notation for operations expressing basic mathematical and computational concepts.
The notation for data flow, control flow, storing, 
and exception handling looks rather like a programming language,
but it is intepreted as pure mathematical functions
(the interpretation depends on whether exceptions are used).

\paragraph{Monads.}

The types of the mathematical functions used in denotational semantics can be quite complicated.
In 1989, Eugenio Moggi suggested that each feature should be seen as a monad, 
where the elements represent computations of values in arbitrary domains \cite{Moggi1989AVPL};
moreover, the required domains could be defined modularly, by applying a series of monad constructors.
Monads have a binary operation for composing a computation with a function that takes its computed value,%
\footnote{See \cite{Mosses2011VDM} for discussion of earlier uses
of similar operations in denotational semantics.}
corresponding to the funcon $\NAME{give}(X, Y)$,
and a unary operation for giving a value as the result of a computation (not needed with funcons).
Each monad constructor adds further structure to the domain of computations,
together with associated operations.
For example, the monad constructor for stores in a domain $S$ makes computations of values in $T$
take an argument in $S$
and return both a value in $T$ and a store in $S$.
The associated operations are $\textit{lookup}(l)$,
to return the value at location $l$ in the argument store and the unchanged store, 
and $\textit{update}(l,v)$, to return a null value and a store where the value at $l$ is $v$.
The funcons $\NAME{assigned}(\VAR{Var})$ and $\NAME{assign}(\VAR{Var}, V)$ correspond to
(a typed variant of) the operations defined by the store monad constructor.
Other funcons closely correspond to the operations associated with monad constructors
for a wide range of notions of computation. 
Monad constructors also need to lift definitions of operations to the resulting domains,
which is non-trivial.
The notation for monad constructors and operations varies
(also between functional programming languages and proof assistants that support monads).

\paragraph{Abstract Semantic Algebras and Action Semantics.}

In a series of papers in the 1980s, the present author proposed various sets of combinators,
together with algebraic laws that they were supposed to obey,
giving so-called \emph{abstract semantic algebras}.
The elements of abstract semantic algebras were intended to have a clear operational interpretation;
they were referred to as `actions' from 1985.

The action notation used in the action semantics framework \cite{Mosses1992AS,Mosses1996TPAS}
was developed in collaboration with David Watt \cite{Mosses1987UAS}.
It was defined \cite[App.~C]{Mosses1992AS} using a novel (but non-modular)
variant of structural operational semantics,
and use of action semantics was supported by tools implemented in the \textsc{Asf+Sdf} Meta-Environment 
\cite{Brand2006AE,Deursen1996ASD}.

Action notation involves actions, data, and yielders.
The performance of an action represents information processing behaviour. 
Yielders used in actions may access, but not change, the current information.
The evaluation of a yielder always results in a data entity.
Many funcons correspond closely to the combinators of action notation.
The crucial difference is that action notation could not be extended with new features,
due to the non-modularity of its operational definition.
The development of modular structural operational semantics \cite{MSOS}
was directly motivated by the aim of making the definition of action notation extensible,
and avoiding reduction of the many facets of action behaviour
to pure functions in monads \cite{Wansbrough1997MMAS}.


\section{Conclusion} 
\label{sec:conclusion}

The \textsc{PLanCompS} project has defined the behaviour of a substantial collection of funcons,
and illustrated translation of functional and imperative language constructs to funcons \cite{TAOSD,CBS-beta}.
It has also developed their theoretical foundations \cite{FOSSACS2013}.
Specifying languages by translation to funcons appears to be significantly less effort
than with other frameworks.
Funcon definitions and translations have been validated by testing, using generated interpreters;
web pages and PDFs are generated from the same source files,
with hyperlinks from names to definitions to support browsing and navigation.

Much remains to be done.
Current and future work includes:
completion and release of the initial collection of funcons and further tool support;
demonstration of scaling up to translation of a major language such as C$\sharp$;
improvement of the definitions of funcons for multithreading;
defining the static semantics of funcons;
defining funcons for expressing static semantics of language constructs;
proving algebraic laws for funcons;
and investigating whether funcons can be used also for specifying 
the semantics of declarative and domain-specific programming languages.
\textsc{PLanCompS} welcomes new participants who would like to contribute to the development of funcons!

\subsubsection{Acknowledgements.}

Helpful comments on a previous version were provided by
Thomas van Binsbergen, Cliff Jones, Neil Sculthorpe, members of the PL Group at Delft, and the anonymous reviewers.
Thanks to the track organisers Klaus Havelund and Bernhard Steffen
for extra space for the appendices.

The initial development of funcons was supported by an EPSRC grant
to Swansea University for the \textsc{PLanCompS} project (EP/I032495/1).
The author is now an emeritus at Swansea, and a visitor at Delft University of Technology.


\appendix
\input{appendix}


\bibliographystyle{splncs04}
\bibliography{funcons}

\end{document}

%% file: appendix.tex
\section{Data}
\label{app:data}

\subsection{Datatypes}

\subsubsection{Primitive values.}

Conceptually, primitive values are atomic, and not formed from other values.
For large (or infinite) types of primitive values, however,
it is infeasible to declare a separate constant for each value.
So in practice, funcons used to construct primitive values usually take other values as arguments.

\begin{itemize}

\item 
$\NAME{booleans}$ are the values $\NAME{true}$, $\NAME{false}$;
funcons corresponding to the usual Boolean operations are defined.

\item
$\NAME{integers}$ is the built-in type of unbounded integers,
with funcons for the usual mathematical operations.
Funcons corresponding to associative binary operations are extended to arbitrary numbers of arguments.
Subtypes include $\NAME{naturals}$ and $\NAME{bounded}(M, N)$;
compositions with casts to such subtypes correspond to partial operations representing computer arithmetic.

\item
$\NAME{floats}$ is the built-in type of IEEE floating point numbers,
with funcons for the required operations.

\item
$\NAME{characters}$ is the built-in type of all \textsc{Unicode} characters.
Its subtypes include $\NAME{ascii-characters}$ and $\NAME{iso-latin-1-characters}$.
Its funcons incude the UTF-8, UTF-16, and UTF-32 encodings of characters as byte sequences.

\item
$\NAME{null-type}$ has the single value $\NAME{null-value}$, alias $\NAME{null}$.
\end{itemize}

\subsubsection{Composite values.}

Conceptually, composite values are constructed from finite sequences of argument values.
The types of composite values include parametrised algebraic data types, with a generic representation.
Various algebraic datatypes are defined, and new ones can be introduced.
Composite values include also built-in parametrised types of sets, maps, multi-sets, and graphs.

\paragraph{Algebraic datatypes.}

\begin{itemize}

\item
$\NAME{datatype-values}$
are generic representations for all algebraic datatype values.

\item
$\NAME{tuples}( T_1, \cdots, T_n )$
are grouped sequences of values of the specified types.

\item
$\NAME{lists}(T)$
are grouped sequences of values of type $T$, with the usual operations;
$\NAME{strings}$ are lists of characters.

\item
$\NAME{vectors}(T)$
are grouped sequences of values of type $T$, accessed by index.

\item
$\NAME{trees}(T)$
are finite, with values of type $T$ at nodes and leaves.

\item
$\NAME{references}(T)$
are values that refer to values of type $T$.

\item
$\NAME{pointers}(T)$
are references to values of type $T$ or $\NAME{pointer-null}$.

\item
$\NAME{records}(T)$ 
are unordered aggregate values, indexed by identifiers.

\item
$\NAME{variants}(T)$
are pairs of identifiers and values of type $T$.

\item
$\NAME{classes}$
are collections of features, allowing multiple superclasses, used to classify objects.

\item
$\NAME{objects}$ are classified collections of features.

\item
$\NAME{bit-vectors} ( N )$ has instantiations for $\NAME{bits}$ and $\NAME{bytes}$.

\end{itemize}

\paragraph{Built-in datatypes.}

\begin{itemize}

\item
$\NAME{sets}(\VAR{GT})$
are finite sets of ground values of type $\VAR{GT}$.

\item
$\NAME{maps}(\VAR{GT}, T^?)$
are finite maps from type $\VAR{GT}$ to type $T^?$.

\item
$\NAME{multisets}(\VAR{GT})$
are finite multisets of ground values of type $\VAR{GT}$.

\item
$\NAME{directed-graphs}(\VAR{GT})$
have values of type $\VAR{GT}$ as vertices.

\end{itemize}

\noindent
See \cite{CBS-beta} for funcons that operate on the above types of values.

\subsection{Abstractions}

\subsubsection{Generic Abstractions.}

These non-ground values are used for constructing thunks, functions, and patterns.
An abstraction body of computation type $T \TO T'$ may refer to a given value of type $T$,
and compute values of type $T'$.

\begin{itemize}

\item
$\NAME{abstractions}(CT)$ are procedural abstractions of computation type~$CT$.

\item
$\NAME{abstraction}(X)$
constructs an abstraction with dynamic bindings.

\item
$\NAME{closure} ( X )$
computes an abstraction with static bindings. 

\item
$\NAME{enact} ( A )$
evaluates the body of the abstraction $A$.

\end{itemize}

\subsubsection{Thunks.}

The abstraction body of a thunk does not reference a given value.

\begin{itemize}

\item
$\NAME{thunks} ( T )$
are constructed from abstractions with bodies of type~$( \ ) \TO T'$.

\item
$\NAME{thunk}( A )$ constructs a thunk from the abstraction $A$.

\item
$\NAME{force}( V )$ enacts the abstraction of the thunk $V$.

\end{itemize}

\subsubsection{Functions.}

The abstraction body of a function may reference a given value.

\begin{itemize}

\item
$\NAME{functions}( T, T' )$
are constructed from abstractions with bodies of type~$T \TO T'$.

\item
$\NAME{function} ( \NAME{abstraction} ( X ) )$
constructs a function with dynamic bindings.

\item
$\NAME{function} ( \NAME{closure}( X ) )$
computes a function with static bindings.

\item
$\NAME{apply} ( F, V )$
gives the value $V$ to the body of the abstraction of function $F$.

\item
$\NAME{supply}( F, V )$
determines the argument value of a function application,
but returns a thunk that defers evaluating the body of the function.

\item
$\NAME{compose}( F_2, F_1 )$
returns the function that first applies $F_1$ 
then $F_2$.

\item
$\NAME{curry} ( F )$
takes a function $F$ that takes a pair of arguments,
and returns the corresponding `curried' function.

\item
$\NAME{uncurry}( F )$
takes a curried function $F$
and returns a function that takes a pair of arguments.

\item
$\NAME{partial-apply} ( F, V )$
takes a function $F$ that takes a pair of arguments, 
and determines the first argument, returning a function of the second argument.

\end{itemize}

\subsubsection{Patterns.}

The abstractions of patterns match a given value.

\begin{itemize}

\item
$\NAME{patterns}$ are constructed from abstractions
with bodies of computation type $\NAME{values} \TO \NAME{environments}$.

\item
$\NAME{pattern}(A)$
constructs patterns from abstractions~$A$.

\item
$\NAME{match}(X, \NAME{pattern}(A))$
enacts the abstraction~$A$, giving it the value of~$X$.

\end{itemize}

\section{Flow of Control}
\label{app:controlflow}

\begin{itemize}

\item 
$\NAME{left-to-right} (  \cdots )$
evaluates its arguments sequentially, and concatenates the computed value sequences.
Composing it with a funcon having pre-computed arguments prevents interleaving; e.g.,
$\NAME{integer-add} ( \NAME{left-to-right} (  X, Y ) )$
always executes $X$ before $Y$.

$\NAME{right-to-left} (  \cdots )$
is analogous.

$\NAME{interleave} (  \cdots )$
evaluates its arguments in any order, possibly with interleaving,
and concatenates the computed value sequences.

\item
$\NAME{sequential} (  X, \cdots )$ 
executes the command $X$, then any remaining arguments,
evaluating to the same value(s) as the last argument.

\item
$\NAME{effect} (  \cdots )$ 
interleaves the evaluations of its arguments, discarding their computed values,
and gives $\NAME{null-value}$.

\item
$\NAME{choice} (  Y,  \cdots )$
selects one of its arguments, then evaluates it.

\item
$\NAME{if-true-else} (  B,  X, Y )$ 
evaluates $B$ to a Boolean value,
then evaluates either $X$ or $Y$
(which have to compute values of the same type).

\item
$\NAME{while-true} (  B, X )$ 
evaluates $B$ to a Boolean value,
then either executes $X$ (which has to correspond to a command) and iterates, or terminates.

\end{itemize}

\section{Flow of Data}
\label{app:dataflow}

\begin{itemize}

\item
$\NAME{given}$ evaluates to the current value of the auxiliary entity $\NAME{given-value}$.

\item
$\NAME{give}(  X, Y )$ 
evaluates $X$. 
It then executes $Y$ with the value of $X$ as the value of the auxiliary entity $\NAME{given-value}$.

\item
$\NAME{left-to-right-map} (  F,  V^* )$ 
evaluates $F$ for each value in the sequence $V^*$ in the same order,
computing the sequence of resulting values.

$\NAME{interleave-map} (  F, V^* )$ allows interleaving of the evaluations.

\item
$\NAME{left-to-right-repeat}(  F, M, N )$
evaluates $F$ for each integer from $M$ up to $N$ sequentially,
computing the sequence of resulting values.

$\NAME{interleave-repeat} (  F, M, N )$ allows interleaving of the evaluations.

\item
$\NAME{left-to-right-filter} (  P, V^* )$
evaluates $P$ for each value in $V^*$,
computing the sequence of argument values for which the value of $P$ is true.

$\NAME{interleave-filter} (  P, V^* )$ allows interleaving of the evaluations.

\item
$\NAME{fold-left} ( F,  A, V^* )$
reduces a sequence $V^*$ to a single value by folding it
from the left, using $A$ as the initial accumulator value.

$\NAME{fold-right} ( F,  A, V^* )$ is analogous.

\end{itemize}

For any list $L$, the funcon term $\NAME{list-elements}(L)$ evaluates 
to the sequence $V^*$ of elements in $L$,
and $\NAME{list}(V^*)$ reconstructs $L$.
Composition with these funcons allows the above funcons on sequences to be used with lists;
similarly for vectors, sets, multisets, and the datatype of maps. 

\section{Name Binding}
\label{app:binding}

\begin{itemize}

\item
$\NAME{bind-value} (  I, X )$
computes the singleton environment mapping~$I$ to the value computed by~$X$.

\item
$\NAME{unbind} (  I )$
computes an environment that hides the binding of~$I$.

\item
$\NAME{bound-value} (  I )$
computes the value to which $I$ is currently bound
(possibly recursively, via a link), if any, and otherwise fails.

\item
$\NAME{scope} ( D, X )$
first evaluates~$D$ to compute an environment~$\rho$.
It then extends the auxiliary environment entity with~$\rho$ for the execution of~$X$.

\item
$\NAME{closed} ( X )$
prevents references to non-local bindings while evaluating~$X$.

\item
$\NAME{accumulate} (  D_1,  D_2 )$
first evaluates~$D_1$, to compute an environment~$\rho_1$.
It then extends the auxiliary environment entity by~$\rho_1$
for the evaluation of~$D_2$,
to compute an environment~$\rho_2$.
The result is $\rho_1$ extended by $\rho_2$.

\item
$\NAME{collateral} ( D_1, \cdots )$
evaluates its arguments to compute environments.
It returns their union as result, failing if their domains are not pairwise disjoint.

\item
$\NAME{bind-recursively} ( I, X )$ makes $\NAME{bind-value} ( I, X )$
recursive.
It first computes a singleton environment $\rho$ mapping~$I$ to a fresh link $L$.
It then extends the auxiliary environment entity by~$\rho$ for the execution of~$X$,
to compute a value~$V$.
Finally, it sets $L$ to refer to~$V$,
and gives~$\rho$ as the computed result.

\item
$\NAME{recursive}( \VAR{SI},  D )$
makes~$D$ recursive on the identifiers in the set~$\VAR{SI}$.
It first computes an environment $\rho$ mapping all~$I$ in~$\VAR{SI}$ to fresh links.
It then extends the auxiliary environment entity by~$\rho$ for the execution of~$D$,
to compute an environment~$\rho'$.
Finally, it sets the link for each~$I$ to refer to the value of~$I$ in~$\rho'$,
and gives~$\rho'$ as the computed result.

\end{itemize}

\section{Imperative Variables}
\label{app:storing}

\begin{itemize}

\item
$\NAME{variables}$
is the type of all simple variables.

\item
$\NAME{allocate-variable} ( T )$
constructs a simple variable for storing values of type $T$
in a location not in the current store.

\item
$\NAME{recycle-variables}( \VAR{Var}, \cdots )$
removes locations allocated to variables from the current store.

\item
$\NAME{initialise-variable}( \VAR{Var}, V )$
assigns $V$ as the initial value of $\VAR{Var}$.

\item
$\NAME{allocate-initialised-variable}( T, V )$
is a composition of $\NAME{allocate-variable} ( T )$ and
$\NAME{initialise-variable}( \_\,, V )$.

\item
$\NAME{assign}( \VAR{Var}, V )$
stores $V$ at the location of $\VAR{Var}$
when the type contains $V$.

\item
$\NAME{assigned} ( \VAR{Var} )$
gives the value last assigned to $\VAR{Var}$.

\item
$\NAME{current-value} ( V )$
gives the same result as $\NAME{assigned} ( V )$ when $V$ is a simple variable,
otherwise $V$.

\item
$\NAME{un-assign} ( \VAR{Var} )$
makes $\VAR{Var}$ uninitialised.

\item
$\NAME{structural-assign} ( V_1, V_2 )$
assigns to all the simple variables in $V_1$
the corresponding values in $V_2$,
provided that the structure and all non-variable values in $V_1$
match the structure and corresponding values of $V_2$.

\item
$\NAME{structural-assigned} ( V )$
computes $V$ with all simple variables replaced by their assigned values.
When $V$ is a simple variable or a value with no component variables, 
$\NAME{structural-assigned} ( V )$ gives the same result as 
$\NAME{current-value} ( V )$.

\end{itemize}

\section{Abrupt Termination}
\label{app:abrupting}

\begin{itemize}

\item
$\NAME{abrupt} ( V )$
terminates abruptly for reason $V$.

\item
$\NAME{handle-abrupt}( X, Y )$
first executes~$X$.
If~$X$ terminates normally, $Y$ is ignored.
If~$X$ terminates abruptly for any reason,
$Y$ is executed, with the reason as the given value.

\item
$\NAME{finally} ( X, Y )$
first executes~$X$.
On normal or abrupt termination of~$X$, it executes~$Y$.
If~$Y$ terminates normally,
its computed value is ignored, and the funcon terminates in the same way as~$X$;
otherwise it terminates in the same way as~$Y$.

\item
$\NAME{fail}$
abruptly terminates for reason $\NAME{failed}$.

\item
$\NAME{else} ( X_1, X_2, \cdots )$
executes the arguments in turn until either some $X_i$ does \emph{not} fail,
or all arguments $X_i$ have been executed.
The last argument executed determines the result.

$\NAME{else-choice} ( X_1, X_2, \cdots )$
is similar, but executes the arguments sequentially in any order.

\item
$\NAME{check-true} ( X )$
terminates normally if the value computed by $X$ is $\NAME{true}$,
and fails if it is $\NAME{false}$.

\item
$\NAME{checked} ( X )$
fails when $X$ computes the empty sequence of values $(  \  )$,
representing that a value has not been computed.
It otherwise computes the same as $X$.

\item
$\NAME{throw} ( V )$
abruptly terminates for reason $\NAME{thrown} ( V )$.

$\NAME{handle-thrown}( X, Y )$
handles abrupt termination of $X$
for reason $\NAME{thrown} ( V )$ with $Y$.

$\NAME{handle-recursively}( X, Y )$ 
is similar to $\NAME{handle-thrown} ( X, Y )$,
except that another copy of the handler attempts to handle any values thrown by $Y$.

\item
$\NAME{return} ( V )$ 
abruptly terminates for reason $\NAME{returned} ( V )$.

$\NAME{handle-return} ( X )$ 
evaluates $X$. 
If $X$ either terminates abruptly for reason $\NAME{returned} ( V )$,
or terminates normally with value $V$,
it terminates normally giving $V$.

\item
$\NAME{break}$
abruptly terminates for reason $\NAME{broken}$.

$\NAME{handle-break} ( X )$
terminates normally when $X$ terminates abruptly for reason $\NAME{broken}$.

\item
$\NAME{continue}$
abruptly terminates for reason $\NAME{continued}$.

$\NAME{handle-continue} ( X )$
terminates normally when $X$ terminates abruptly for reason $\NAME{continued}$.

\end{itemize}
Further funcons are provided for expressing delimited continuations \cite{CBS-beta,Sculthorpe2016MSOSDC}.

\section{Communication}
\label{app:interacting}

\begin{itemize}

\item
$\NAME{read}$
inputs a single non-null value from the $\NAME{standard-in}$ entity, and gives it as the result.

\item
$\NAME{print} ( V^* )$
outputs the sequence of values $V^*$ to the $\NAME{standard-out}$ entity. 

\end{itemize}